# DESIGNING A PATTERN LANGUAGE FOR SURVIVING EARTHQUAKES

Tomoki Furukawazono
Shota Seshimo
Daiki Muramatsu
Takashi Iba

Keio University
Endoh 5322
Fujisawa, Kanagawa, 252-0816, Japan
e-mail: zono@sfc.keio.ac.jp

**ABSTRACT**

In this paper, we proposed the Survival Language, a pattern language to support survival when a catastrophic earthquake occurs. This proposal comes from the problem that the tragedies of earthquakes are repeated, because knowledge and wisdom on how to prepare for an earthquake and what to do during an earthquake have not been passed down sufficiently. This paper presented the four patterns of the Survival Language: "Daily Use of Reserves," "Life over Furniture," "Evacuation Initiator," and "Kick Signal." In addition, we described that the Survival Language is created and used by collaborations because a pattern language has been a tool for collaboration since it was presented by Christopher Alexander.

**INTRODUCTION**

The objective of this paper is to propose the Survival Language, a pattern language (Alexander 1979; Alexander et. al. 1977) to support actions for survival when a catastrophic earthquake occurs. This proposal comes from the problem that the tragedies of earthquakes are repeated even in Japan, where people have experienced numerous earthquakes, because knowledge and wisdom on how to prepare for an earthquake and what to do during an earthquake have not been passed down sufficiently.

In Japan, infrastructures for earthquakes have been developed from the Great Hanshin-Awaji earthquake of 1995. For example, the Cabinet Office of Japan has been developing an integrated disaster management information system, such as the Earthquake Disaster Information System, Real Damage Analysis System by Artificial Satellite, and Disaster Information Sharing Platform (Cabinet Office, Government of Japan). In addition, from the Great East Japan earthquake of 2011, the Japanese government has been pursuing the maintenance of information and communication technology, such as the decision and deployment of the "Basic Policy and Action Plan for Building IT Disaster-Management Lifeline." (IT Strategic Headquarters 2012)

While such maintenance of infrastructure in the macro level is being pursued, guidance concerning specific actions is still insufficient. In the Great Hanshin-Awaji earthquake of 1995, many suffered from their lack of emergency preparation. Yet many similar instances occurred in the Great East Japan earthquake of 2011. These instances reveal that the knowledge and wisdom concerning specific actions for survival during and immediately after earthquakes have not been passed on sufficiently.

Therefore, this paper proposes the Survival Language as a methodology that supports people to make immediate decisions possible when an earthquake hits and to integrate earthquake preparation in their daily lives. The Survival Language intends to design one's immediate actions when an earthquake occurs, because it is critical to accumulate one's knowledge and combine them in such circumstances. Another intention is to constantly remind one about the significance of earthquake preparation, because one's awareness of catastrophic earthquakes that seldom occur tends to become weary gradually.

Although the pattern language proposed in this paper is written based on catastrophic earthquakes in Japan (the Great Hanshin earthquake of 1995, the Great East Japan earthquake of 2011, and so on), such earthquakes have occurred around the world, such as the 1994 Northridge earthquake in USA, the 2011 Virginia earthquake in USA, the 1960 Valdiva earthquake in Chile, the 2009 L'Aquila earthquake in Italy, the 2008 Sichuan earthquake in China, the 2004 Indian Ocean earthquake, the 2011 Christchurch earthquake in New Zealand, the 2011 Haiti earthquake, and the 2011 Kütahya earthquake in Turkey. By writing a pattern language from the knowledge and wisdom learned from the past earthquakes in Japan, the Survival Language seeks to support actions for survival through future catastrophic earthquakes around the world.

## A PATTERN LANGUAGE FOR SURVIVING EARTHQUAKES

A pattern language is a set of "patterns" which each scribe out the complex relationships of a person's knowledge, especially of those which are tacit and usually embedded deeply into the person's mind and actions (Alexander 1979; Alexander et. al. 1977). Through a mining process, in this case through interviews, these knowledge are verbalized and scribed out. These kinds of knowledge often come in the form of a solution to a complex problem. Thus, these patterns are written in a rather strict format, containing the "context" in which a "problem" occurs, and the "solution" to this problem is the knowledge which has value to be written out. This set of information is grouped together as a "pattern", and then is given a name. The complex knowledge then can be referred to by the pattern name, making communication about the idea to occur easily. Having the patterns in their mind would also allow users to cut out and recognize patterns out of otherwise unnoticed sequence of events. There are multiple patterns in a pattern language, and the users of the language choose and combine patterns out of the language to design their own actions.

The Survival Language is organized into one whole structure in order to achieve survival when an earthquake occurs. Figure 1 shows the overview of the whole structure of this language. This structure is formed by three categories: "Designing Preparation", "Designing Emergency Action", and "Designing Life after Quake." "Designing Preparation" consists of patterns to use during one's daily life before an earthquake occurs. "Designing Emergency Action" consists of patterns to use immediately when an earthquake actually happens. "Designing Life after Quake" consists of patterns to use within 72 hours after the earthquake (the mortality rate significantly rises after this time period.)

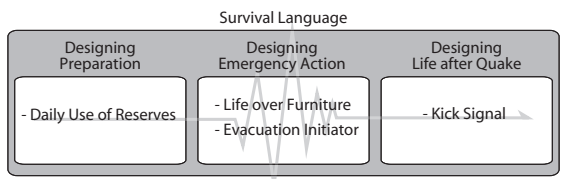

Figure 1: Overview of the Survival Language

Although this language is planned to be consisted of a few dozen patterns, this paper will show the four patterns out of the set. We propose: "Daily Use of Reserves" from the "Designing Preparation" group, "Life over Furniture" and "Evacuation Initiator" from the "Designing Emergency Actions" group, and "Kick Signal" from the "Designing Life after Quake." In addition to these four patterns, a few dozen patterns are currently being written: "1981 Line" from the "Designing Preparation" group, "Armadillo Pose" and "Evacuation before Fire Fighting" from the "Designing Emergency Action" group, and "Shrine Shelter" from the "Designing Life after Quake." The Survival Language seeks to support survival in a catastrophic earthquake by combining patterns from each of these three categories.

## DESCRIPTIVE FORMAT OF THE SURVIVAL LANGUAGE

The Survival Language seeks to support immediate decisions when an earthquake strikes, and to recall earthquake safety measures even in ordinary moments of daily life. Therefore the format of the patterns in the Survival Language must be written in a manner which is clear and easy to recall. Thus the Context of the Survival Language is written in a style which would allow the person to recall contents of the pattern instantly when an earthquake occurs and quick action is necessary. In addition, the writing style of the Context is designed so that one can quickly search through one's mind. For instance, instead of "When you are with multiple people, an earthquake occurred," the context reads: "An earthquake occurred, and there are people around you." Furthermore, the Problem and Solution is also written in a simple manner which can be completely memorized, and be easily recalled. The Forces and Actions describe the Problem and Solution more specifically.

Each pattern is written in the same form: Pattern Name, Introductory Sentence, Picture, Context, Problem, Force, Solution, Action, Consequence and the Case. The Pattern Name is the attractive and memorable names that could be used as building blocks for the thinking process and as a vocabulary for communicating for survival in an earthquakes; the Introductory Sentence and Picture are introductory parts of the patterns to present the pattern livelier; the Context describes the conditions for when survivors should apply the pattern. The Problem describes difficulties that often occur in the context, and the Forces are unavoidable laws that make the problem difficult to solve; the Solution describes how to solve the problem, with Actions offering concrete methods to put the solution into practice; and Consequence describes the result of applying the pattern. Finally, the Case demonstrates various episodes from people who have experienced major earthquakes of Japan in the past. Most of these episodes (Atsumi 2007; the Cabinet Office of Japan) are published on the web pages of the Cabinet Office, the Japanese Government , and of local governments in Japan.

## THE SURVIVAL LANGUAGE AND COLLABORATION

### A Pattern Language and Collaboration

The Survival Language is created and used by collaborations because a pattern language has been a tool for collaboration since a pattern language for architectures was presented by Christopher Alexander.

A pattern language has a close relationship with collaboration. First of all, a pattern language has been a tool for collaboration. Many people consider that a pattern language is something like a manual which they should follow in creating architectures and that they can construct more better buildings than any other. However, such idea is not true. Generally, a pattern language is used as a common language for users and architects to design architecture. Some patterns are used and others are not. Such use of the pattern depends on the collaboration of the users and architects.

Furthermore, a pattern language itself is written through the collaboration of the architects and the users. Discussing the purposes of the buildings and types of patterns needed, the users and architects make each pattern. Thus, a pattern language is an outcome of such collaboration.

The six fundamental principles of Alexander's experiment at Oregon University are clearly proven since a pattern language is a tool for and result of collaboration (Alexander et. al. 1975). We would like to introduce those six principles below (Alexander et. al. op.cit., p.5-6).

---

**1. The principle of organic order.**
Planning and construction will be guided by a process which allows the whole to emerge gradually from local acts.

**2. The principle of participation.**
All decisions about what to build, and how to build it, will be in the hands of the users.

**3. The principle of piecemeal growth.**
The construction undertaken in each budgetary period will be weighed overwhelmingly towards small projects.

**4. The principle of patterns.**
All design and construction will be guided by a collection of communally adopted planning principles called patterns.

**5. The principle of diagnosis.**
The well being of the whole will be protected by an annual diagnosis which explains, in detail, which spaces are alive and which ones dead, at any given moment in the history of the community.

**6. The principle of coordination.**
Finally, the slow emergence of organic order in the whole will be assured by a funding process which regulates the stream of individual projects put forward by users.

---

The principle that most exemplifies collaboration is the principle of participation. Other principles cannot exist without collaboration. Alexander and his team conducted their research following such principles. Creating a pattern language with the users, faculties, and students at Oregon University, Alexander designed architecture based on that pattern language.

A Pattern Language has been made through and has been the outcome of collaboration. Similarly, the Survival Language can also be a result and used as a tool for collaboration.

### The Survival Language and Collaboration

The Survival Language is created and used by collaborations as other pattern languages is. We would like to explain the two collaborative points of the Survival Language: Collaborative Creations and Collaborative Uses.

*The Survival Language and Collaborative Creations*

The four patterns of the Survival Language in this paper are created by a collaborative process (Gloor 2010).

The creators of the Survival Language for communication are mainly the authors of this paper. The members communicate through information communication technologies such as Mailing List, Wiki, SKYPE, Instant Messenger, Dropbox and twitter. The procedure of creating the Survival Language consists of the following five phases (Figure 2): (1) Pattern Mining, (2) Pattern Prototyping, (3) Pattern Writing, (4) Language Organizing, and (5) Catalogue Editing (Iba et. al. 2010). This process is not strictly sequential, and allows the creators to go back and forth amongst these phases if necessary.

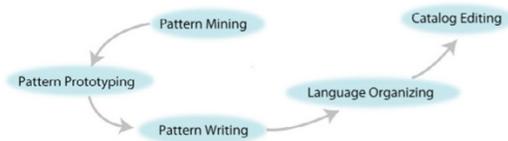

*Figure 2: Five Phases of Creating Patterns*

In addition, each phase is open to the public, with a Collaborative Innovation Network. Members of The COIN are mainly the other members of the Takashi Iba Laboratory. These members contribute to the creation of the Survival Language through participating in discussions in each phase. Through these discussions, each member provides his/her own individual knowledge derived from experiences or expertise. The creators integrate the COIN's contributions into creating patterns. For example, the name of the pattern: "Life over Furniture" is named by a member of the COIN.

*The Survival Language and Collaborative Uses*

People can help each other, using this pattern language when an earthquake erupt. When an earthquake strikes, it will take time for the government to help the sufferers. Therefore, the earthquake victims need to collaborate in helping each other before the government's aid reaches them. For example, using the pattern called "Kick Signal" in the Survival Language, earthquake victims who are closed under the buildings can notify their danger to others. This will increase the possibility of people who realizes the "Signal" and help the sufferers.

When an earthquake strikes, victims need to collaborate in helping each other. In this situation, the Survival Language is useful as a tool for collaboration.

## CONCLUSION

This paper proposed the Survival Language, which aims to support people with their survival when a catastrophic earthquake occurs, and presented four patterns out of the set. These four patterns are just a part of the Survival Language, and we will present more patterns in the future. The pattern language introduced in this paper can be classified as patterns to support human actions in a broad sense. We are planning to convene a workshop concerning disaster safety with the Survival Language, to support people's actual safety measures. In the future, to make survival possible when catastrophic earthquakes occur, we would like to develop and improve this pattern language even further.

## ACKNOWLEDGEMENTS

First of all, we should like to express our deepest gratitude to Ko Matsuzuka for his writing support. We also would like to thank Kaori Harasawa for her personal qualities and her friendly support. Without their support, this work would not be possible. We also want to thank Aya Matsumoto, Megumi Kadotani and Taichi Isaku for brushing up this paper.

## APPENDIX

Here are four patterns from the Survival Language, A Pattern Language for Surviving Earthquakes: "Daily Use of Reserves," "Life over Furniture," "Evacuation Initiator," and "Kick Signal."

No.1

# Daily Use of Reserves

*Use and replenish reserves on a daily basis, and you are safe in urgent times.*

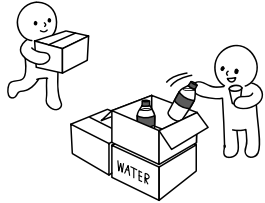

**To prepare for a catastrophic earthquake, you stockpiled emergency food and supplies.**

▼In this context

**You forget to refresh your reserves, and the food and supplies will have passed their expiration date when about to use.**

Many people actually have the feeling that they need to prepare for a catastrophic earthquake, and stockpile emergency food and supplies. However, often times these reserves pass their expiration dates. A collective reason accounts for this problem. First of all, catastrophic earthquakes do not occur frequently. It may occur tomorrow, or it may never occur in your lifetime. If you plan to use reserves only when a catastrophic earthquake occurs, there is a possibility that you will not use them for a very long time, thus reaching their expiration date. In addition, because reserves are generally stored in places away from the usual living space in the house, your awareness for them fade, and you tend to forget the exact expiration date. Even if you remember the expiration date exactly, buying all reserves every time they expire can be very expensive, and you hesitate to restore your reserves.

▼Therefore

**Use your stockpiled food and supplies on a daily basis. Then, replenish the same amount you used.**

To prevent reserves from expiring, it is necessary to apply a system where you use the reserves daily, and replenish the amount used. For example, when storing drinking water, it is impractical to store several boxes of water only for emergency purposes, apart from regular use. Rather, buy some extra cases than usual (e.g. three cases), and store them for emergency, but also use them on a daily basis in order of earlier expiration dates. Then, buy the amount you used. For instance, after using up one case and you have two left, buy one new case. That way, you will always have at least two fresh cases as reserves.

▼Consequently

**Emergency food and supplies constantly stay fresh.**

By using and replenishing reserves on a daily basis, you will constantly have fresh food and supplies as reserves. In a sense, you are applying the same system of how grocery stores can use their storage as a source for emergency supplies when an earthquake happens into a more personal level. Even if the magnitude of the earthquake is small, social infrastructure such as mailing and shipping services may fall into chaos. In such situations, because product supply stalls or buyouts may occur, obtaining necessary products will become difficult. However, if you already have fresh reserves, you can retain emergency food and supplies even in such chaotic circumstances.

---

*CASE*
In a particular household, the family buys canned or instant food more than the usually amount when they are on discount, in order to stock a month worth of reserves. These reserves are consumed from the oldest on a daily basis, allowing them to never expire. When the family shops at a grocery store, they buy canned or instant food each member enjoys, even if those foods may not stay fresh longer than foods specifically designed for emergency. This allows this family to consume reserves they enjoy on a regular basis. In addition, since these reserves are stored in a place where every member of the family can see (e.g. drawers, shelves, food storages), they can easily check the amount left and the expiration date of their current reserves.

# Life over Furniture

Don't try to support your furniture - run.

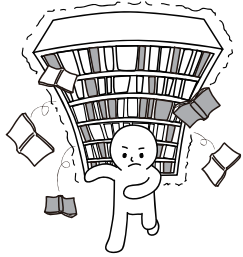

**During an earthquake, furniture around you are shaking, making them likely to collapse or its contents to fall out.**

▼In this context

**You try to support the shaking furniture, but they will consequently collapse on you.**

When an earthquake hits and furniture start to shake, although you may feel danger, you also feel the urge to support the furniture from collapsing, or keep the objects inside from falling. However, when the earthquake intensifies, even light furniture that may look like it could be held down easily will have potential dangers in which objects inside may burst out and hit your body(such as your head.) Furthermore, when the earthquake becomes severe, not only can objects inside burst out of furniture, but also the furniture themselves can fall or slide. If the furniture falls on you, your body will be trapped, and in the worst case, you may lose your life from the pressure.

▼Therefore

**Restrain from holding shaking furniture away from them.**

If you feel an shake, immediately get away from any furniture nearby. Especially look out for tall furniture, or those with breakable objects inside. For example, when the shaking intensifies, dishes can burst out violently from inside dish shelves. Prevent injuring yourself with broken pieces of glass or dishes by getting away from it as much as possible. In addition, books on bookshelves, or heavy objects placed on higher shelves are also similarly hazardous. Your life may be in danger especially if the furniture itself falls on you, so it is critical to go away as fast and as much as you can.

▼Consequently

**You can prevent furniture from collapsing and its contents from falling on yourself.**

By immediately getting away from any furniture nearby, you can protect yourself from collapsing furniture or falling objects. Objects inside furniture or furniture themselves may break, but at least your body will be safe. For instance, with dish shelves, dishes shatter, and the shelf itself may break. However, if you attempt to hold the shelf from falling, glass or dishes may hit you, and you may suffer severe injuries. With bookshelves, valuable books may fall off, and may be destroyed. However, if you attempt to hold the bookshelf, countless books may fall on you, and in worst case scenario, put your life in danger. By restraining from supporting shaking furniture, and immediately going away, you can save your life from danger.

> *CASE*
> On March 20th, 2005, an earthquake struck the Genkainada area located in the northwestern sea of the Fukuoka prefecture, Japan. A family lived on the sixth floor of an apartment in this area. After coming home from shopping, the daughter was editing a photo on her laptop. When the earthquake struck, she heard an abnormal sound, and saw a three-meter high Buddhist altar flying through the air. Japanese Buddhist altars are traditionally very large and heavy, and are not seen as an object that would normally "fly." The daughter did not try to hold down the expensive piece of furniture, but ran out of the house. As a result, even though the Buddhist altar broke and cost six million yen to repair, more importantly, she was not injured, and was safe.

**Evacuation Initiator**

Initiating the run will also save others.

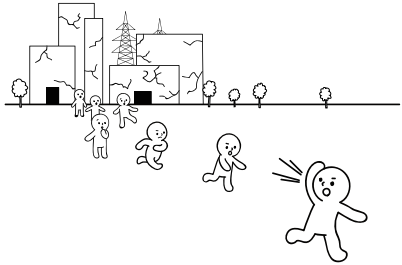

**An earthquake occurred, and there are people around you.**

▼In this context

**You are influenced by people around who do not seem to move, and you cannot evacuate in time.**

When an earthquake happens, the fear from the shake brings the idea of evacuating. However, since neither the magnitude of the earthquake nor its source is known, we tend to hesitate taking the run. When looking around, the people around are also uncertain on what to do, but no one seems to be evacuating. Some people even are saying that there is no need to evacuate in effort to calm the people around down. These conditions are giving you the idea that it is still safe and an evacuation is unneeded, but as aftershocks and additional earthquakes happen, the crowd is faced with even worse danger.

▼Therefore

**Shout out loud to let people around you aware, and take the initiative in the evacuation.**

When an earthquake hits, become an Evacuation Initiator and take the first move to evacuate. When doing so, shout out loudly to tell the people around to move also. Although specifics about the earthquake that just happened is unknown, nor may all people not evacuate with you, taking the first action to trigger others to move to a safe place is top priority. When at school or office, use the stairs and not an elevator to get outside. If outdoors when an earthquake hits, watch out for any falling trees and buildings, and head for a big field far from anything around. A school field, or a large park are good candidates for places to evacuate to.

▼Consequently

**You can evacuate with the people around you.**

By shouting out loud while running, it will take away the qualm from running away alone. It would make the people around aware of the danger, and they could move to save their lives too. Even though it may turn out that an evacuation was not needed afterwards, it will not hurt to move to a safe place just in case when consequences of not evacuating may have brought is imagined.

This solution is not suggesting that you should run instantly without thinking whenever an earthquake occurs. Stop and judge if the current place you are in is safe before running. If it is, stay there. Escaping from the place will only bring up the possibility of injuries. In that case, advise other people to come to your place so they would be safe too.

---

*CASE*
One of the authors, Muramatsu, of this paper was at a skiing site in Yamagata about to ride on a lift when the Tohoku Earthquake of 2011 had hit. At first he had thought that it was only an small earthquake as usual, but the shake seemed to be getting stronger. As he saw the lift platform building shaking, he was convinced that this was going to be a big earthquake. He was still thinking about how much stronger this earthquake was going to get, when his professor Iba, also an author of this paper, said "let's run," and started to head out of the building. This action made Muramatsu realize that this place was dangerous, and he started to run with his professor. After they got out of the building, they heard an announcement from the building cautioning people that the building was old and may collapse. If the two had not taken quick action, the building may have collapsed on top of them.

No.4

**Kick Signal**

If you are trapped, kick your surroundings and signal out an SOS.

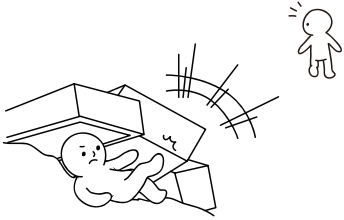

**You are surrounded by fallen debris, and are trapped inside.**

▼In this context

No matter how loudly you shout for help, your voice is muffed by objects around you, and people outside do not notice you. When a catastrophic earthquake occurs, parts of a building or furniture may collapse on you, and you may get trapped inside. It is very difficult to escape from fallen debris by yourself, and you call out for help. However, no matter how loudly you call, your voice will be muffed by fallen debris, and people outside may not notice you. Rather, you lose stamina from the continuous loud shouts you make. In addition, because no one notices your existence even when you shout, you feel hopeless, thinking that you cannot be saved.

**Kick your surroundings, and signal out a banging noise.**

▼Therefore

If parts of a building or furniture fall, and you are trapped under them, kick your surroundings and signal a banging noise outside. For instance, if there is a desk that fell near your leg, kick it as hard as you can, and make a banging noise. Because the vibrations fro this noise passes on to other debris, it is more likely to be noticed. If you sense someone is near, make as loud of a banging noise as you can. In some cases, you may not be able to move your leg. In such conditions, bang objects with your hand, or even your head. You can also use a stick if one is available near your hand. In one way or another, it is important to send out a banging noise from under the debris.

**You can notify people outside about your existence, and your chances of being rescue increases.**

▼Consequently

By kicking your surroundings and making a banging noise, you can signal your existence to people outside. When parts of a building or furniture fall on you, they may restrict bodily movements, or produce large amounts of dust which may choke your eyes, mouth, or nose. Trying to make noises as loud as you can from under debris with your leg or arms, and sometimes even head can be very tough. However, even if you shout for help, the chances of notifying people outside are low. But if you make a banging noise with all of your strength, and the noise passes on outside, and you can signal your existence outside. And when your signal goes outside, your chances of being rescued rises.

---

*CASE*
When the Great Hanshin earthquake struck in 1995, the first floor was crushed by the floor above in the Hagino couple's wooden house. The couple was sleeping on the first floor, when their house collapsed on them. Though fortunately they did not lose their lives in the crash, they were trapped under the debris. Their son, who was sleeping on the second floor, escaped from the fallen house. He called for his parents from outside the debris in order to rescue them, and the couple shouted theirs son's name to answer it. However, the couple's voice was muffed by the fallen debris, and did not reach their son. The son took the wrong assumption that his parents have already evacuated, and he headed for an emergency shelter. Although he realized later that he was wrong and returned, he couldn't tell if they were still alive. However, the noise of the son's return reached inside the fallen house, and the husband used the only appendage he could move: his leg, and kicked a low table up. The noise he made reached outside, and the son was able to realize his parents' survival. Seven hours after the earthquake struck, the couple was safely rescued from their fallen house.